\documentclass[letter]{aa} 
\usepackage{graphicx}
\usepackage{txfonts}
\usepackage{natbib}
\newcommand{\beq}   {\begin{equation}}
\newcommand{\eeq}   {\end{equation}}
\newcommand{\kms}   {km~s$^{-1}$}
\newcommand{\water}   {H$_2$O~}
\def\gs{\mathrel{\raise0.35ex\hbox{$\scriptstyle >$}\kern-0.6em
\lower0.40ex\hbox{{$\scriptstyle \sim$}}}}
\def\ls{\mathrel{\raise0.35ex\hbox{$\scriptstyle <$}\kern-0.6em
\lower0.40ex\hbox{{$\scriptstyle \sim$}}}}

\begin{document}
   \title{Possible magnetic field variability during the 6.7~GHz methanol maser flares of G09.62+0.20}
   \titlerunning{Magnetic field variability in G09.62+0.20}


   \author{W.H.T. Vlemmings\inst{1}\and
           S. Goedhart\inst{2}\and M.J. Gaylard\inst{2} }

   \offprints{WV (wouter@astro.uni-bonn.de)}

    \institute{Argelander Institute for Astronomy, University of Bonn,
     Auf dem H{\"u}gel 71, 53121 Bonn, Germany
         \and
     Hartebeesthoek Radio Astronomy Observatory, P.O. Box 443, Krugersdorp, 1740, South Africa}

   \date{Received 08/04/2009; accepted 29/04/2009}

   \abstract{Polarization of maser emission contains unique
     information on the magnetic field in the densest regions of
     massive star formation.}{Recently, the magnetic field induced
     Zeeman splitting was measured for the strongest known 6.7~GHz
     methanol maser, which arises in the massive star forming region
     G09.62+0.20. This maser is one of a handful of periodically flaring
     methanol masers. Magnetic field measurements can possibly provide
     insights into the elusive mechanism responsible for this
     periodicity.}{The 100-m Effelsberg telescope was used to monitor
     the 6.7~GHz methanol masers of G09.62+0.20, in weekly intervals,
     for just over a two month period during which one of the maser
     flares occurred.}{With the exception of a two week period during
     the peak of the maser flare, we measure a constant magnetic field
     of $B_{\rm ||}\approx11\pm2$~mG in the two strongest maser
     components of G09.62+0.20 that are separated by over 200~AU. In
     the two week period that coincides exactly with the peak of the
     maser flare of the strongest maser feature, we measure a sharp
     decrease and possible reversal of the Zeeman splitting.}{While
     the two phenomena are clearly related, the Zeeman splitting
     decrease only occurs near the flare maximum. Intrinsic magnetic
     field variability is thus unlikely to be the reason for the maser
     variability. The exact cause of both variabilities is still
     unclear, but it could be related to either background
     amplification of polarized emission or the presence of a massive
     protostar with a close-by companion. However, the variability of
     the splitting between the right- and left-circular polarizations
     could also be caused by non-Zeeman effects related to the
     radiative transfer of polarized maser emission. In that case we
     can put limits on the magnetic field orientation and the maser
     saturation level. }

   \maketitle
%

\section{Introduction}

Polarization observations of astrophysical masers provide important
insights into the magnetic field properties of, among others, the
dense regions surrounding massive protostars \citep[e.g.][and
references therein]{Vlemmings07a}. In these regions, magnetic fields
may play an crucial role in, e.g., suppressing fragmentation, altering
feedback processes and stabilizing accretion disks. Linear
polarization observations of maser emission can reveal the magnetic
field morphology, while observations of the circular polarization,
generated due to Zeeman splitting, can be used to measure the
line-of-sight magnetic field strength. Especially for the
non-paramagnetic molecules such as SiO, \water and methanol, maser
polarization also depends on intrinsic maser properties that determine
the maser saturation level, such as brightness temperature, beaming
angle and the rate of maser stimulated emission. Thus, polarization
observations can, in addition to the magnetic field strength and
structure, also provide constraints on maser properties that are
otherwise hard to determine \citep[e.g.][]{Vlemmings06}. Circular
polarization, or Zeeman splitting observations have been mostly
focused on OH and \water masers \citep[e.g.][]{Hutawarakorn99,
  Sarma01, Bartkiewicz05, Vlemmings06}. However, recent
  observations have revealed significant Zeeman splitting of the
  6.7~GHz $5_1-6_0A^+$ methanol transition \citep[][hereafter
  V08]{Vlemmings08a} in a sample of 17 out of 24 of the brightest
  northern methanol maser sources, indicating an average magnetic
  field strength in the maser region of $|B|\sim20$~mG. 

The massive star forming region G09.62+0.20 harbors the strongest
known maser at 6.7~GHz. This maser, and its 12.2~GHz counterpart,
undergoes periodic flares with a period of 244 days
\citep[e.g.][]{Goedhart03, Goedhart04}. At the height of its flare,
the 6.7~GHz maser has been seen to reach a peak flux density of over
7000 Jy~beam$^{-1}$. There is a time-delay of $\sim$25 days between
the flare in the strongest 6.7 GHz feature (hereafter the {\it main}
feature) at $\sim1.2$~\kms and the {\it secondary} feature at
$\sim-0.1$~\kms, which also displays a different flare profile. The
origin of the periodic behavior, however, is still unclear. The
G09.62+0.20 star forming region consists of a complex of H{\sc II}
regions in various evolutionary stages. Its 6.7~GHz methanol masers
have been mapped with the ATCA by \citet{Phillips98}, and the
strongest features are shown to be associated with the hypercompact
H{\sc II} region labeled E by \citet{Garay93}. This region is
speculated to be excited by a B0 star \citep{Hofner96}. VLBA observations of the 12.2~GHz masers during the
course of a flare \citep{Goedhart05} indicate that the maser regions
simply brighten in intensity, with no change in morphology, implying
that the cause of the flare arises beyond the maser region. It has
been previously speculated that the periodicity could arise from
either the background H{\sc II} region, or an infrared pump
source. The monitoring observations presented here were prompted by
the detection of a possible magnetic field reversal between the main
($B_{\rm ||}\approx -3$~mG) and secondary ($B_{\rm ||}\approx 9$~mG)
maser features in V08.

\begin{figure}[t!]
   \resizebox{0.90\hsize}{!}{\includegraphics{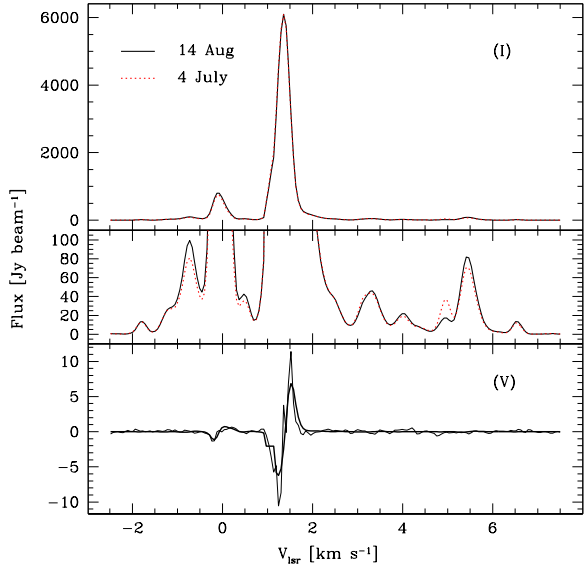}} \hfill
\caption[specs]{The 6.7~GHz methanol maser total intensity (I) spectrum of G09.62+0.20 for two observing epochs. The peak flux at July 4 is scaled to the flux at Aug 14 2008 (top). The middle panel highlights the many weaker maser features of this region. The bottom panel shows the circular polarization spectrum (V) for the observations at Aug 14 2008. The thick solid line is the best fit fractional total intensity derivative.}
\label{Fig:spec}
\end{figure}

\section{Observations and error analysis}
\label{obs}

The 6668.519~MHz ($5_1 - 6_0A^+$) methanol maser line of the massive
star forming region G09.62+0.20 was monitored weekly over a period of
slightly more than 2 months (11 sessions from 2008 June 14 to August
20) using the 5~cm primary focus receiver of the 100-m
Effelsberg\footnote{The 100-m telescope at Effelsberg is operated by
  the Max-Planck-Institut f{\"u}r Radioastronomie (MPIfR) on behalf of
  the Max-Planck-Gesellschaft (MPG)} telescope. The observing dates
were chosen to encompass the expected flare events of the main and
secondary variable maser features. Unfortunately a setup problem made
two sessions (June 14 and July 31) unusable. In addition to
G09.62+0.20, we observed G23.01-0.41 as a consistency check. As in the
previous Effelsberg methanol maser Zeeman splitting observations
(V08), the data were taken in position switch mode with a 2 minute
cycle time. Data were collected using the fast Fourier transform
spectrometer using two spectral windows, corresponding to the right-
and left-circular polarizations (RCP and LCP). The spectral windows of
$20$~MHz were divided in $16384$ spectral channels, resulting in a
$\sim0.055$~\kms channel spacing and were centered on the local
standard of rest (LSR) source velocities. The data were reduced as
described in V08, with amplitude calibration performed on 3C286. The
spectrum of G09.62+0.20 for two Effelsberg observational epochs is
shown in Fig.~\ref{Fig:spec}, along with the circular polarization
spectrum for one of the epohcs. As seen in the figure, the shape
  of each maser feature is identical at each epoch. However, the
  velocity of the maser peak of the main feature shifts similarly in
  RCP and LCP by up to $\sim10$~m~s$^{-1}$. This is likely caused by a
  delay of the flare of weaker maser features in the wings of the main
  feature (Goedhart et al., in prep.).

The target sources G09.62+0.20 and G23.01-0.41 were observed for 10
and 6 minutes respectively. Due to changing conditions, the rms noise
levels varied over the different monitoring sessions. Additionally,
the LCP rms noise level was between 50\% and 300\% larger than the RCP
rms noise level. The LCP rms noise, which varied for the 10~min
observations of G09.62+0.20 between 100 and 350~mJy, was thus the
limiting factor to our Zeeman splitting determinations. Unfortunately,
as the rms noise level was a factor of $\sim2$ increased over that in
the V08 observations, we were only able to detect significant Zeeman
splitting in our consistency check source G23.01-0.41 at 4 of the
epochs. Within the errors, these epochs were consistent with the V08
observations.

Synchronous observations of G09.62+0.20, as part of a program
to monitor the variable methanol maser sources
\citep{Goedhart05}, were carried out using the 26m telescope
at Hartebeesthoek Radio Astronomy Observatory (HartRAO). Full sampling
of the flare was not possible due to scheduling constraints. Flux
calibrations were done using continuum drift scans across Hydra A and
3C123 and a further check was made using the methanol maser
G351.42+0.64 as a comparison source. Comparison with the HartRAO
measurements enables us to estimate the absolute flux errors to be
less than $10\%$.

\begin{figure}[t!]
   \resizebox{0.90\hsize}{!}{\includegraphics{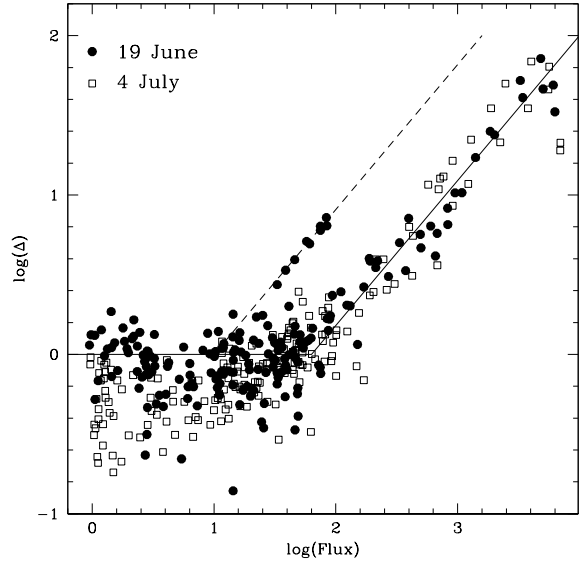}} \hfill
   \caption[rmsinc]{The ratio $\Delta$ of the true and predicted
     channel rms noise as a function of channel flux for two
       epochs. The solid line indicates a two-component description
     of the rms noise increase due to dynamic range limits of the
     spectrometer. The dashed line indicates the anomalous increased
     rms noise for the maser features at $V_{\rm LSR}=5.4$~\kms seen
     in the first epoch.}
\label{Fig:rms}
\end{figure}

As our observations were aimed at detecting possible small
variabilities of the Zeeman splitting, we performed an additional
analysis of the channel rms noise in the spectrum of G09.62+0.20. To
determine the increase of channel rms noise as a function of maser
flux, we used the 5 individual 2 minute scans of two of the epochs to
determine the channel rms in those channels that contained significant
maser emission. The individual maser spectra were normalized to the
peak flux of the combined scans to minimize the effect of intrinsic
maser variability. We then calculate $\Delta_{\rm i}={\rm
  rms_i}/{\sigma}$ for each channel $i$, where $\sigma$ is the rms
noise value for the emission free channels. Fig.~\ref{Fig:rms} shows
$\Delta$ as a function of the maser flux in each of the channels with
$>5\sigma$ maser emission. We find that for all of the maser
  features, with one exception, $\Delta$ stays approximately constant
  within a factor of $\sim3$ up to a maser flux of
  $\sim50$~Jy~beam$^{-1}$ after which it increases with
  $\Delta\propto({\rm Flux})^{0.9}$.  During the first epoch, only
the maser feature at $V_{\rm LSR}=5.4$~\kms did not follow this
relation and already deviated from the expected rms noise level when
its flux became $>20$~Jy~beam$^{-1}$. As this did not occur during the
last epoch, this is possibly due to weak narrowband interference.   Since the noise characteristic is similar for both polarizations and
  all epochs, the rms error increase is unlikely to be due to receiver
  saturation and would be unable to cause a systematic shift between
  RCP and LCP. However, the analysis does imply that the errors on
  Zeeman splitting determination in V08 should be increased. For the
majority of the masers in V08, this increase is less than a factor of
$\sim5$. However, for the masers of G09.62+0.20 the errors need to be
increased with a factor of $40$ and $8$ for the main and secondary
maser features respectively.

\section{Variability of RCP-LCP frequency splitting}
\subsection{The case of G09.62+0.20}

\begin{figure}[t!]
   \resizebox{\hsize}{!}{\includegraphics{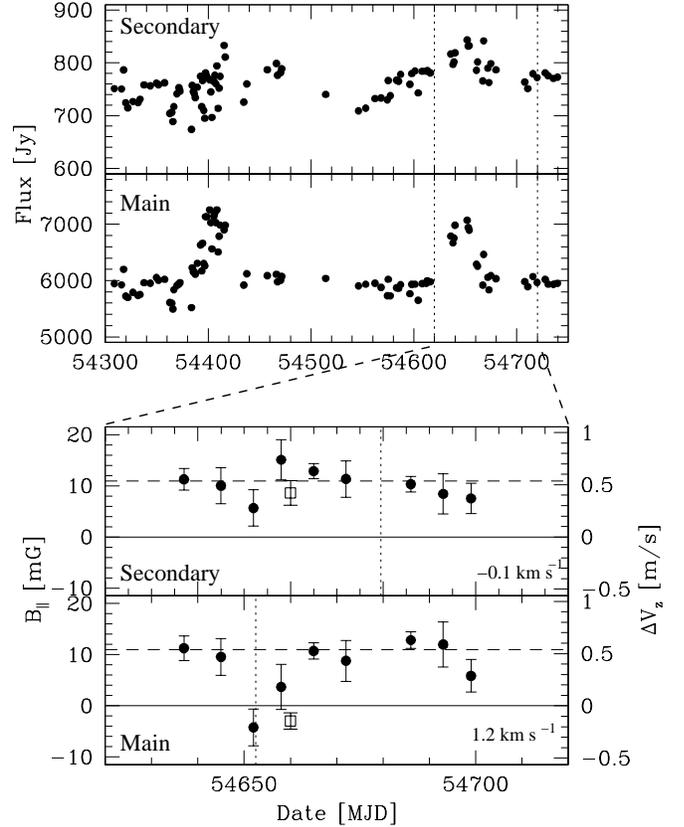}} \hfill
\caption[varplot]{(bottom panels) The Zeeman splitting $\Delta V_Z$ (in m~s$^{-1}$) and derived line-of-sight magnetic field strength $B_{\rm ||}$ (in mG) for the two strongest 6.7~GHz methanol maser features of G09.62+0.20 at the 9 successful monitoring epochs (filled dots). The previous observations of Nov 11th 2007 (open square; V08) has been folded into the new observations.
The vertical short dashed lines indicate the predicted date of the emission peak. The horizontal dashed lines indicate the weighted average magnetic field strength (see text). (top panels) HartRAO telescope observations of the two maser features for two flaring periods. The period of the Effelsberg observations is indicated by the vertical short dashed lines.}
\label{Fig:var}
\end{figure}

We determined the Zeeman splitting of the two brightest 6.7~GHz
methanol maser features of G09.62+0.20 using the RCP-LCP
cross-correlation method as described in V08. Fig.~\ref{Fig:var} shows
the line-of-sight magnetic field strength $B_{\rm ||}$ for the two
maser features at each epoch as derived from the Zeeman splitting
$\Delta V_Z$, using $0.049$~\kms~G$^{-1}$ as the best estimate for the
Zeeman splitting coefficient. In addition to the monitoring epochs,
the figure also includes the folded in observations presented in V08. We find that the line-of-sight
magnetic field strength in the secondary maser feature is stable
during the observations, with an error weighted average magnetic field
of $B_{\rm ||}=10.9\pm2.3$~mG. In contrast, the Zeeman splitting of
the main maser feature sharply decreases at the
time the maser flare reaches peak flux. Monitoring at weekly intervals
has proven to be too coarse to put strong constraints on the length of
the period with decreased $\Delta V_Z$. However, while the typical
duration of a flare is up to two months, the decrease in Zeeman
splitting lasts only for an approximate two week period around the
peak of the flare. Determining a error weighted average magnetic field
for the main maser feature using the 7 epochs on either side of
the two week period with decreased Zeeman splitting, we find $B_{\rm
  ||}=11.0\pm2.2$~mG.  Thus, the magnetic field strength is remarkably
similar on both masers, even though both features are
separated by more than 200~AU \citep{Goedhart05}.

Fig.~\ref{Fig:var} thus indicates that, when the flare of the main
maser feature reaches its peak flux, the Zeeman splitting decreases
significantly and potentially even changes sign. As the observations
of V08 that also revealed a much lower (and possibly reversed)
$\Delta V_Z$ were taken close to the peak of the previously flaring
period, this behavior appears to repeat itself regularly. No significant Zeeman
splitting decrease is seen for the secondary maser feature. The
observations with the HartRAO telescope showed that the flare of this
feature, while typically approximately 25 days after the flare of the
main feature, was much more irregular and reached its peak at
approximately the same time as the main feature. However, throughout
the Effelsberg observations its flux variations were only $\sim10\%$
while that of the primary feature was over $20\%$.
 
\subsection{Possible origin of the observed variability}

The cause for the sudden decrease in Zeeman splitting during the maser
flare is unknown. We can confidently rule out instrumental effects as
the reason for the decrease of $\Delta V_Z$ of the main maser feature
for a number of reasons. First of all, no significant corresponding
decrease is found for the secondary maser feature at the same
epochs. Furthermore, the quite constant Zeeman splitting measured
before and after the peak, and the fact that the negative Zeeman
splitting was also found during the maser flare 8 months earlier,
point to the stability of the instrumental setup as well as the
robustness of the data reduction and analysis method.

The observed effect is thus likely intrinsic to the source. The
measured decrease in $\Delta V_Z$ however, starts after the maser flux
has already entered the flaring stage. Thus, while the measured Zeeman
splitting variation is related to the maser flare, it is unlikely that
the flare itself is caused by changes in the magnetic field. Here we
describe three possible scenarios that could give rise to the observed
effect.


{\it Background amplification:} As the observed Zeeman splitting is generated by the average magnetic
field throughout the entire maser region, the most straightforward
cause of a drop in observed Zeeman splitting is the superposition of
two maser regions with opposite magnetic fields. As described in
\citet[e.g.][]{Boboltz98}, maser flares can often be attributed to the
chance alignment of maser regions, when a foreground maser amplifies
the emission from a background maser. Alternatively, instead of a
maser, the background source could be a strongly polarized continuum
source. In this case, polarization of the seed emission can cancel
out any circular polarization generated within the maser. However,
while this simple model also naturally explains the maser flare,
the origin of the observed periodicity remains unclear. 


{\it Intrinsic magnetic field variability:} A second option is that the observations reveal an actual change in
the magnetic field within the maser region. Various mechanisms can
produce a change in magnetic field. It has been suggested that the
periodic maser flares are due to changes in the maser pumping resulting
from a binary interaction between a massive protostar and companion
\citep{Goedhart03}. Interactions between the magnetospheres of the two
companions could cause periodic behavior of the magnetic field, as
has been observed in young low-mass binary systems
\citep[e.g.][]{Massi08}. However, typical magnetic reconnection events
do not last long enough to explain the approximate two weeks duration
of the magnetic field variability. It is also unclear if the magnetic
interaction would be noticeable in the methanol maser region at
several hundred AU distance from the protostars. Still, an embedded
binary will also give rise to other complex interactions besides that
of the magnetic field, such as those between possible accretion disks
and outflows. These interaction will be imprinted onto the
observed magnetic field. Thus, if the observed magnetic field
variability is truly due to intrinsic changes of the magnetic field,
any possible explanation of the periodic maser flares will also need
to take into account the behavior of the magnetic field.


{\it Maser radiative transfer:} Alternatively, we are observing a combination of Zeeman and non-Zeeman
effects, competing when the maser is at its brightest. Non-Zeeman
effects were briefly discussed in V08. Specifically the effect where
the axis of symmetry for the molecular quantum states rotates when the
maser saturates, and the rate of stimulated emission $R$ becomes
larger than the Zeeman frequency shift $g\Omega$, bears further
investigation. An unfortunate error was introduced in the calculation
of $g\Omega$ presented in V08, with the true $g\Omega$ being smaller
than presented there. For a typical magnetic field strength $B$, in
the dense maser region, of order $10$~mG, $g\Omega\approx13$~s$^{-1}$,
approximately three times larger than the rate of stimulated emission
$R\sim4$~s$^{-1}$ for the most saturated 6.7~GHz methanol masers with
a maser brightness temperature $T_b\sim10^{12}$~K~sr. This assumes
  a typical maser beaming angle $\Delta\Omega=10^{-2}$~sr, which
  decreases rapidly when the maser saturates. Thus, for most typical
methanol masers with $T_b$ of order a few times $10^{10}$~K,
$R<0.1$~s$^{-1}<<g\Omega$ and little or no intensity dependent
polarization is generated. This is supported by the fact that no
relation between maser intensity and Zeeman splitting was found in
V08. However, the brightest masers of G09.62+0.20 could have an $R$
that approaches or even becomes larger than $g\Omega$, raising the
possibility of intensity dependent circular polarization mimicking the
Zeeman splitting between the RCP and LCP spectra. The generation of
circular polarization due to this effect has been investigated for a
$J=2-1$ transition by \citet{NW90}. Although the effect is likely
smaller for the transition of the 6.7~GHz methanol masers involving
higher angular momentum states, it is found that the sign of the
circular polarization, and consequently the sign of the splitting
between RCP and LCP, is opposite from that generated by the regular
Zeeman effect when the angle between the magnetic field and the maser
line of sight $\theta$ obeys $\sin^2\theta<2/3$. Observationally, this
implies that with increasing maser brightness temperature and
consequently $R$, the observed splitting between RCP and LCP spectra
decreases. As the average Zeeman splitting of the two maser features,
with fluxes different by an order of magnitude, are identical, the
non-Zeeman effect apparently only becomes important when the flare
nears its maximum flux. This indicates the masers of G09.62+0.20
approach complete saturation at its peak. It will be possible to test
the hypothesis of the non-Zeeman interpretation of the observed
splitting variability by simultaneously observing the maser linear
polarization, as similar considerations predict an intensity
dependence of fractional linear polarization and polarization angle.

\section{Conclusions}

We have presented Effelsberg 100-m telescope monitoring observations
of the RCP-LCP frequency splitting of the periodically flaring 6.7~GHz
methanol masers of the massive star forming region G09.62+0.20. A
significant decrease in Zeeman splitting and thus possibly magnetic
field strength is detected on the strongest maser feature during a two
week period surrounding the flare maximum. Besides this decrease, a
remarkable constant $B_{\rm ||}$ of $\sim11$~mG is detected for the
two brightest maser features separated by over 200~AU. The cause for
the decrease in measured Zeeman splitting is still unclear but it is
either related to the mechanism that causes the periodic maser flaring
or a result of a non-Zeeman effect when the maser saturation level
becomes significant.

Thus, G09.62+9.20, with the relative predictability of the flares,
varying levels of intensities amongst maser features and correlated
variability in RCP-LCP frequency splitting, is an ideal natural
laboratory to test theories relating to both massive star formation
and maser physics.

\begin{acknowledgements}
  WV acknowledges support by the \emph{Deut\-sche
    For\-schungs\-ge\-mein\-schaft} through the Emmy Noether Research
  grant VL 61/3-1. WV also thanks Alex Lazarian and Bill Watson for kindly
  answering a number of his questions.
\end{acknowledgements}

\end{document}